\documentstyle[aps,twocolumn,epsfig]{revtex}
\begin{document}
\draft

\title{Instabilities of wave function monopoles in Bose-Einstein condensates}

\author{Th. Busch, J. R. Anglin}

\address{Institut f\"ur Theoretische Physik, 
Universit\"at Innsbruck, 
A--6020 Innsbruck, AUSTRIA}

\maketitle

\begin{abstract}
We present analytic and numerical results for a class of 
monopole solutions to the 
two-component Gross-Pitaevski equation for a two-species Bose 
condensate in an effectively two-dimensional trap.  We exhibit dynamical 
instabilities involving vortex production as one species pours through 
another, from which we conclude that the sub-optical sharpness
of potentials exerted by matter waves makes condensates ideal tools for
manipulating condensates.  We also show that there are two equally valid but
drastically different hydrodynamic descriptions of a two-component 
condensate, and illustrate how different phenomena may appear simpler in each.
\end{abstract}


\date{\today}

\narrowtext

Monopoles are particle-like field configurations with which one can
associate a topological charge. As perhaps the most obvious way of
making a collective particle out of a condensate field
\cite{Science95,Bradley,Davis}, they are a natural goal for
condensate state engineering.  In this Letter we discuss a specific case
$D=2$ of a class of $D$-dimensional monopoles, previously introduced
in general and examined for $D=1$\cite{MEDS}.  Although the notion of
topological charge (in $D=2$, the `winding number') is useful in
identifying and classifying these particle-like configurations, in our
case this charge is not guaranteed to be conserved.  Indeed the
structures we discuss are quite prone to dynamical instabilities.
They are nevertheless also robust, in that their instabilities
typically do not destroy them, but rather distort them, in ways which
are both intrinsically interesting, and illustrative of some basic
features of multi-component superfluid dynamics. 

Our paper is organized as follows.  After presenting our concept of a
`wave function monopole', we consider a one-dimensional `ring
monopole' which bears much the same useful analogy to the 2D
monopole as the one-dimensional persistent current does to the vortex
(and which also merits serious consideration in its own right).  In
this context we briefly discuss the consequences of having different
scattering lengths among the two atomic species.  We then present and
analyze the results of numerical solutions to the two-component
Gross-Pitaevski equation.  The specific cases we examine are chosen,
mainly for theoretical simplicity, from among a vast range of
possibilities, and while we have also investigated enough other cases
to confirm that the results we present are not qualitatively altered
by slight changes in parameters, we are not insisting that our precise
examples will be optimal laboratory subjects in the near future.  We
do identify some basic features of a class of structures that are
realistic targets for experiments, and so our paper concludes with a
few remarks of experimental relevance.

\vbox{\noindent\parbox[c]{\linewidth}{~~\parfillskip=0pt 
The basic concept of a monopole in $D$ dimensions is sketched in
Fig.~1 (for $D=2$). The arrows represent an\vspace*{-2 pt} }
\hbox{\noindent\parbox[b]{.4\linewidth}{\begin{center}
\epsfig{file=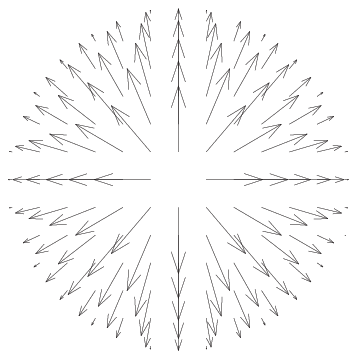,width=.8\linewidth} {FIG.~1. Monopole.\\}
\end{center}}
\noindent\parbox[b]{.585\linewidth}{\parfillskip=0pt  order
parameter having as many components as the dimensionality of the space
in which the monopole exists. There is much freedom in
the actual identification of this order parameter.  For instance, in
two dimensions its magnitude and direction could represent the modulus
and phase of the macroscopic wave function}}}\noindent of a one-species
condensate; this `monopole' is the well-known superfluid vortex.  In
the case we now consider, however, the two dimensions of the arrows
instead represent the wave functions of each component of a
two-species condensate, in a configuration in which both components
happen to be real: $(\psi_1,\psi_2) = f(r) (\cos\theta, \sin\theta)$,
where $r,\theta$ are the usual polar co-ordinates\cite{spinmono}.

As with the vortex, the modulus of the order parameter is the density
of the condensate (here, the total density of both components), so
that the `particle-like' core of the monopole is in fact a local
minimum of density: a void or bubble, maintained by destructive
interference of matter waves.  As with the vortex, however, it is
instructive first to avoid the core $r\to0$ of the monopole, and
consider only its behaviour at large $r$, where the radial dimension
becomes unimportant, leaving the effectively
one-dimensional problem of a two-species condensate on a circle of fixed
radius $R$. So we replace $f(r)\to 1$, and examine the `ring monopole'
$(\psi_1,\psi_2)$ as a stationary
solution to the one-dimensional Gross-Pitaevski equation
\begin{equation}\label{ringGPE}
i\dot\psi_j = -{1\over2R^2}{\partial^2\over\partial\theta^2}\psi_j
 +g(|\psi_1|^2+|\psi_2|^2-\mu)\psi_j\;,
\end{equation}
where $\mu = 1-(2g R^2)^{-1}$ provides $\dot{\psi}_j=0$. 

For the ring monopole we can obtain analytically the 
Bogoliubov spectrum of 
perturbations, $\psi_j \to \psi_j + 
\epsilon\phi(\theta,t)$.
Working to linear order in $\epsilon$ yields the modes
\begin{eqnarray}\label{ringbog}
\left(\matrix{\phi_{1k}^\pm\cr \phi_{2k}^\pm}\right)&=& 
\cos(\Omega_{k\pm}t)\left(\matrix{X_{1k}^\pm \cr X_{2k}^\pm}\right)\nonumber\\
&&\qquad -i{2R^2\Omega_{k\pm}\over k(k+2)}\sin(\Omega_{k\pm}t)
        \left(\matrix{Y_{1k}^\pm \cr Y_{2k}^\pm}\right)\\
\left(\matrix{X_{1k}^\pm \cr X_{2k}^\pm}\right)
&=&\left(\matrix{(k-2)\cos(k-1)\theta 
+ C_{k\pm}\cos(k+1)\theta\cr -(k-2)\sin(k-1)\theta+
        C_{k\pm}\sin(k+1)\theta}\right)
        \nonumber\\
\left(\matrix{Y_{1k}^\pm \cr Y_{2k}^\pm}\right)
&=&\left(\matrix{(k+2)\cos(k-1)\theta 
+ C_{k\pm}\cos(k+1)\theta\cr -(k+2)\sin(k-1)\theta+
                C_{k\pm}\sin(k+1)\theta}\right)\;.\nonumber
\end{eqnarray}
The $\pm$ index distinguishes acoustic and optical branches, in which
the two species' density perturbations are respectively in and out of
phase:
\begin{eqnarray}\label{optacoust}
C_{k\pm} &=& 2+{2k^2\over g R^2} \pm k\sqrt{1+8(g R^2)^{-1}
+4k^2(g R^2)^{-2}}\nonumber\\
{\Omega_{k\pm}^2\over k^2} &=& {g\over2R^2}+{k^2+4\over4R^4}\\
&\pm&\sqrt{\left({g\over2R^2}+{k^2+4\over4R^4}\right)^2
        -{k^2-4\over4 R^6}\left(g+{k^2-4\over4R^2}\right)}
                \;.\nonumber
\end{eqnarray}
Note that rotating these modes by 
$\theta\to(\theta + \pi/2)$ produces an independent set of modes, with the
same frequencies; every frequency is thus two-fold degenerate.  

The modes $k=0$ (for which the $\pm$ branches co-incide) and 
$k\pm=2-$ provide the four zero modes due to the
model's U(2) symmetry (which already includes spatial rotation).  All
the other modes are positive frequency excitations, except for the two
$k\pm=1-$ modes; and for $g R^2 > 3/4$, these become dynamical
instabilities.  Infinitesimal excitation of these modes then grows
exponentially until it becomes finite, and the Bogoliubov theory is
inadequate.  To follow the evolution into this regime, we solve the
Gross-Pitaevski equation numerically, using the split operator
method as described in
\cite{MEDS}.  It turns out that the finite perturbation does not mix
with other modes to produce irreversible (long revival time) decay,
but grows to a maximum size, then shrinks back; and the cycle repeats.  This
indicates that the two unstable modes form an isolated subsystem of
two degrees of freedom, with a `Mexican hat' potential.\cite{breakdown}

\begin{center}
 \begin{figure}
 \epsfig{file=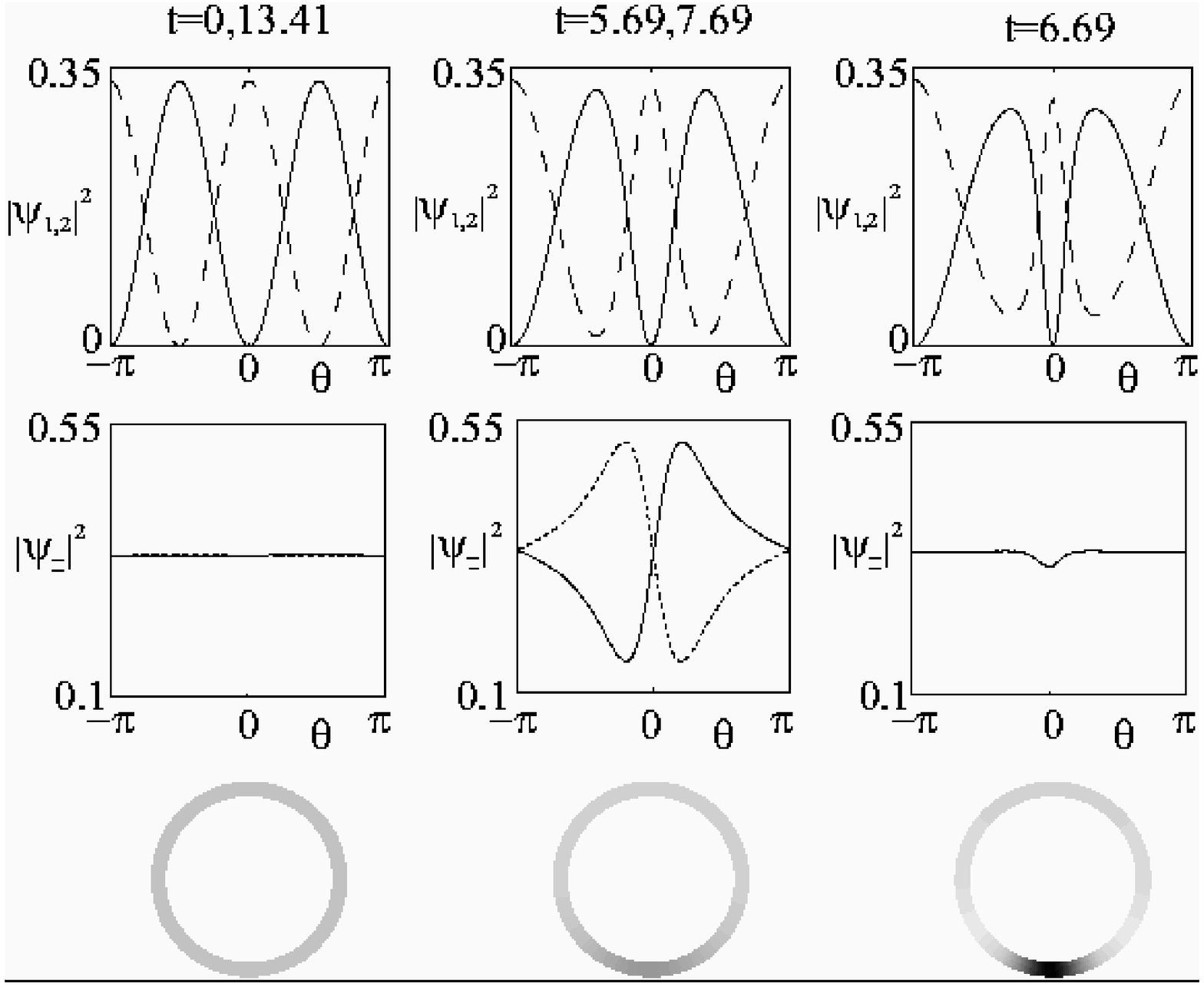,width=.8\linewidth}
 \caption{Dynamical instability of the ring monopole, for 
$gR^2=300/\pi$. Five times are shown, the configurations at the last two
times being precise revivals of those at the first two. The top three
plots show $|\psi_1|^2$ (broken) and $|\psi_2|^2$ (solid); the middle three
show $|\psi_+|^2$ and $|\psi_-|^2$ for the same configurations.  (In the left
middle plot the small initial perturbation can just be seen.)  
The three rings at bottom are overhead views of the $|\psi_1|^2 + |\psi_2|^2$,
with darker shading for lower total density.  The location of the dark spot
is determined by the initial conditions, but if a small random potential 
is added on the ring, over many cycles the system performs what might be called
`superfluid roulette'.} \end{figure}
\end{center}

Before going on to examine instabilities of monopoles in two dimensions, we 
pause here to address an issue of generalization, which will lead to the 
development of some concepts useful for understanding two-dimensional 
phenomena.
In (\ref{ringGPE}) it is assumed that the three scattering lengths and
two particle numbers in the problem are all equal; this is very close
to the case for $^{87}$Rb \cite{2comp}. In fact the simple solution we
examine here can be generalized to encompass all positive scattering
lengths $g_{ij}$.  When separation is favoured ($\det{g}<1$)\cite{Ho},
one can obtain solutions in which our cosine and sine are replaced by
Jacobi elliptic functions.  (Phase-separating monopoles in two dimensions
resemble crossed domain walls, dividing two condensates each with a Josephson 
$\pi$-junction through the monopole core.  They appear to 
have qualitatively similar behaviour to the simplest case with $g_{ij}=g$.)  
When mixing is favoured
($\det{g}>1$), one can also generalize monopoles, but in a somewhat
surprising way.  If one merely changes basis, defining $\psi_\pm =
{1\over\sqrt2}(\psi_1\pm i\psi_2)$, one discovers that our monopole
can also be described as two superimposed, counter-rotating vortices
(SCV).  Changing scattering lengths so as to favour mixing of the two
species $\psi_\pm$ will then stabilize this configuration.

Of course, this change of basis is only a
symmetry of the system when all scattering lengths are equal (and
hence only then is it really true that the monopole and the SCV are
the same object).
Nevertheless the two bases can always be used, and so one has two very
different hydrodynamic pictures which describe the same superfluid
physics. 
This is our first principle, that with two-component
superfluids one has two equally valid hydrodynamic descriptions, and
that one is free to base intuition on either of them.  As will be
apparent below, the differences between them can be dramatic.

We now return to the case where all scattering lengths are equal, and
investigate the monopole in two dimensions.  We note that $(\psi_1,\psi_2)= 
f(r)(\cos\theta,\sin\theta)$ is a stationary solution to the Gross-Pitaevski 
equation (GPE)
\begin{equation}\label{2dGPE}
i\dot\psi_j = -{1\over2}\nabla^2_{2}\psi_j
 +g (|\psi_1|^2+|\psi_2|^2-\mu)\psi_j + {1\over2}r^2\psi_j\;,
\end{equation}
as long as $f(r)$ satisfies the same nonlinear equation as the modulus of
the two-dimensional vortex in the trap:
\begin{equation}\label{feqn}
f'' + {1\over r}f' +({r^2\over2}-{1\over r^2})f= 2g(f^2-\mu)f\;.
\end{equation}

We have studied such 
monopoles in harmonic traps numerically, using the split-operator
method to solve the two-component GPE in two dimensions, in both real and
imaginary time.  We have examined a wide range of different parameter regimes,
and compared results with grids of varying sizes; and while below we
show displacements and motions which are exactly in one of our grid directions,
we have also checked that the same behaviour occurs at arbitrary angles. 

If a monopole is displaced from the centre of the trap, it begins
to fall back towards the centre; but this does not mean that
it is stable. By considering the linearized theory in imaginary
time, one can show that there are lower energy configurations near 
the monopole,
reached by short range perturbations near the core.  While this
analysis does not suffice to determine whether the instability is
dynamical or merely energetic, numerical study of small amplitude
oscillations of a monopole in a trap indicates that monopoles are
dynamically unstable: the amplitude of small oscillations grows 
exponentially.

\begin{center}
\begin{figure}
\epsfig{file=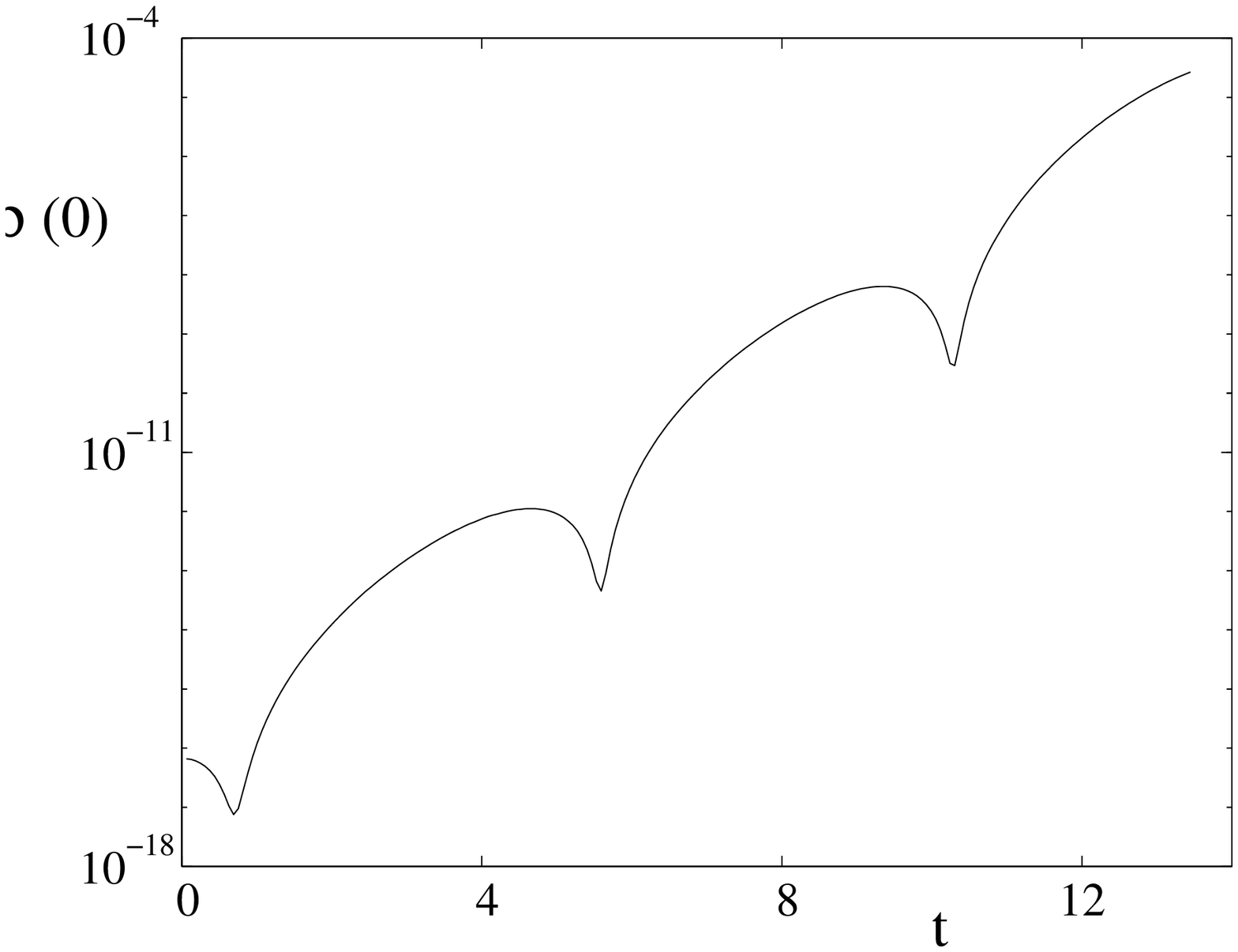,width=.5\linewidth}
\caption{Total density $\rho=|\psi_1|^2+|\psi_2|^2$ in the centre of
a harmonic trap, for a monopole initially displaced very slightly 
from the centre, at $g\int\!d^2x|\psi_j|^2 =2000$.  Note logarithmic vertical
scale. Successive minima do not reach $\rho=0$, because growing speed of the
monopole involves filling in of the core.}
 \end{figure}
\end{center}

The instability, when infinitesimal, is a self-amplifying oscillation
of the monopole back and forth through the centre of the trap; this is at 
least qualitatively similar to the instability on the ring.  As the 
motion grows finite, vortex pairs form.   As can be seen from
Fig.~4, the motion of the monopole involves the pouring of one of the
two species through a narrow channel formed by the other species (which is
comparatively passive throughout).  As shown in Fig.~5, very 
rapid flow develops along the sides of this channel, until vortex pairs 
suddenly nucleate at the maxima of the passive species' density.  Each pair
then separates, into an inner vortex pinned by the passive species, and an 
outer vortex, which gradually drifts away.  Fig.~6 shows this process
as the growth of a pair of branch cuts in
the condensate phase.  It may be quite hard to
directly detect vortex pairs formed in this manner; but their appearance
vividly illustrates a potentially useful fact: 
that sharp-edged potentials (such as most readily
generate vortices) are much more easily formed with matter waves than
with light waves.  
If one wishes to manipulate condensates with high 
resolution forces, therefore, the best tools are probably other condensates.

Despite spawning vortices, an initially
displaced monopole continues its fall towards the centre of the trap with
little apparent impediment.  As (or just before) it reaches the centre, 
however, its motion changes.  In most cases we have investigated, it abruptly
slows (though in some cases, and for reasons which are not clear to us,
it seems instead to accelerate, and then decay rather violently).  Thereafter,
the minimum in total density at the monopole core gradually fills in, until
only a slight depression remains.   
But even at this late stage, a double-lobed 
pattern of the two densities (as in Fig.~4) remains; and in each species 
a phase difference of $\pi$ persists across the monopole, although the 
sharp jump in phase has been smoothed out.  The
instability appears to have saturated, leaving a `smeared'
monopole which appears to be dynamically (although not energetically) stable.  
This eventual stability is much easier to
understand in the SCV
basis. There, all that happens is that the superimposed
counter-rotating vortices slip apart, becoming a pair of reciprocally
pinned vortices: each condensate species has one vortex, and each also
fills a strong potential well formed by the other species' vortex.
(This may be seen as an analogue of the SCV instability on the ring.) 
Once this separation has occurred, the stability of two pinned
vortices is unsurprising. See Fig.~7.

\begin{center}
 \begin{figure} \epsfig{file=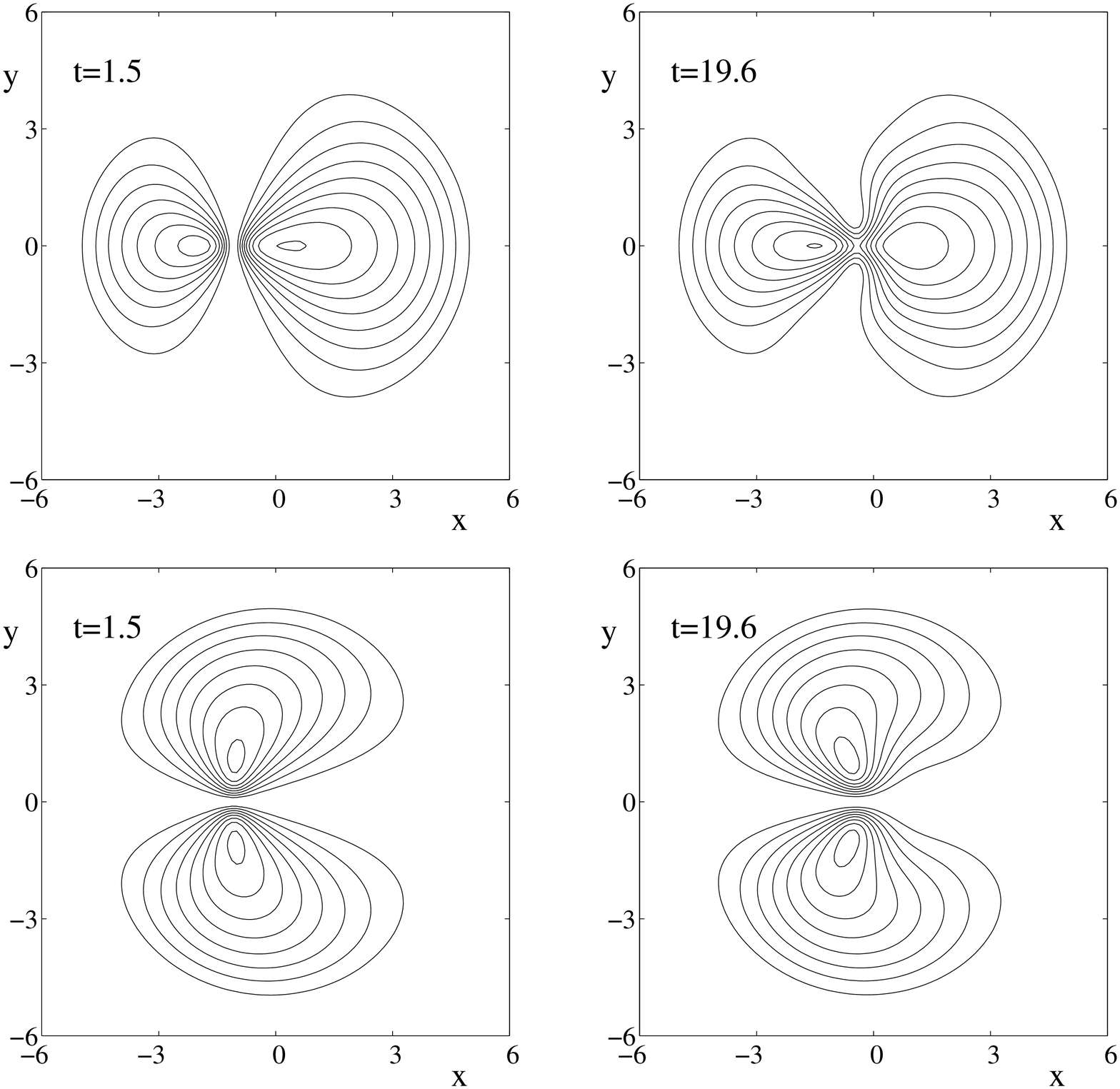, width=.7\linewidth}
 \caption{$|\psi_1|^2$ (top) and $|\psi_2|^2$ for a 
   monopole initially displaced in the direction $\theta = \pi$, at
   early and late times ($g\int\!d^2x|\psi_j|^2 =2000$).  Note
   `throttle' formed by $|\psi_2|^2$.}
\end{figure}
\end{center}

\begin{center}
 \begin{figure} \label{fig:velo} \epsfig{file=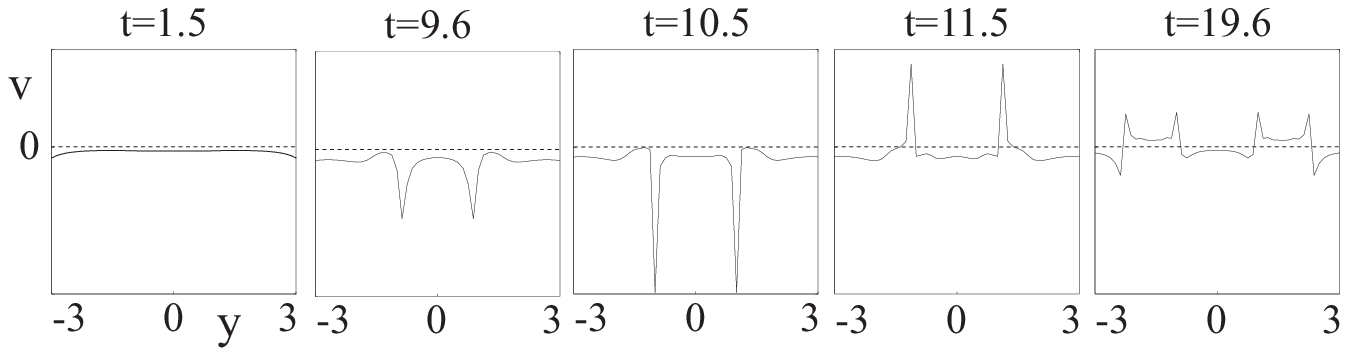,
 width=.95\linewidth} \caption{Superfluid velocity component 
$v_{x1}=\Im(\partial_x\ln\psi_1)$, at various times in the same simulation 
illustrated
in Fig.~4.  Plots are `edge-on' views of the velocity field $v_{x1}(x,y)$, 
looking in the positive x direction: the velocity is very small everywhere 
except near the arc on which $|\psi_1|^2$ vanishes, so the edge-on view is
convenient.)  Initial fluid velocity is negative as the monopole moves
in the positive ($\theta=0$) direction.}
\end{figure}
\end{center}

\begin{center}
 \begin{figure}
 \epsfig{file=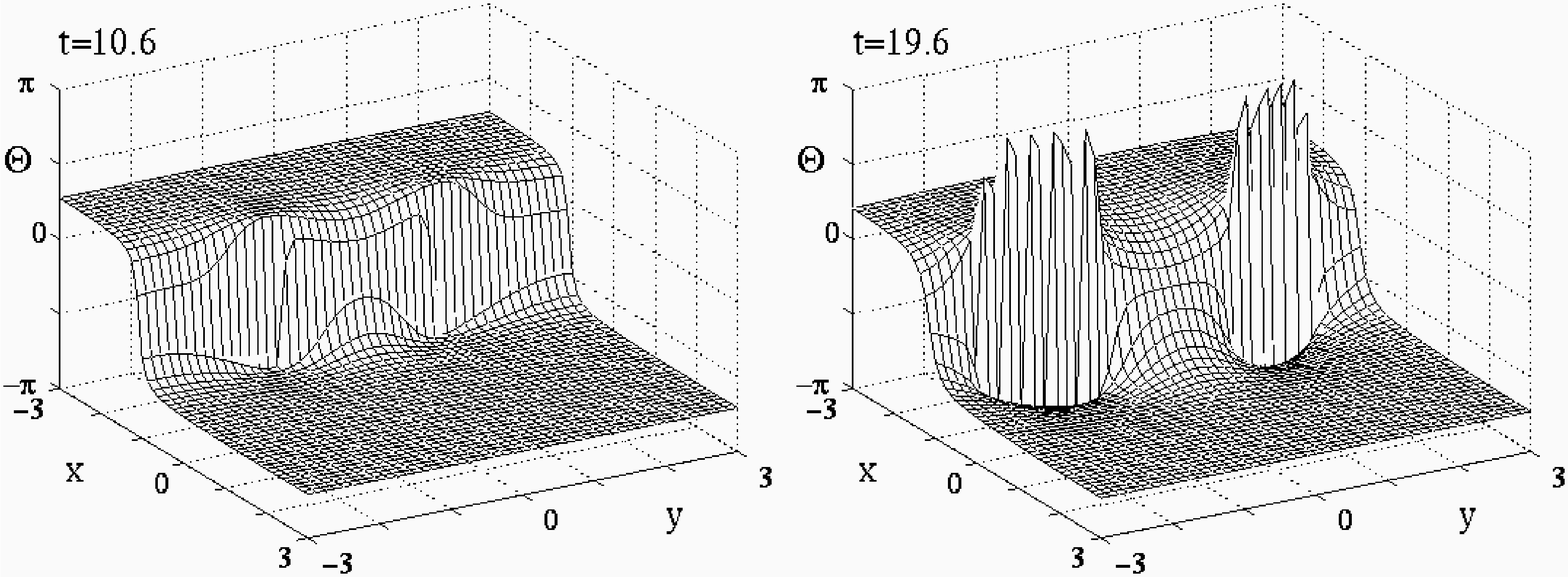, width=.95\linewidth}
 \caption{Phase of $\psi_1(x,y)$ before and after appearance of vortex pairs.
The branch cuts are indeed jumps of $2\pi$.}
 \end{figure}
\end{center}

\begin{center}
 \begin{figure} \epsfig{file=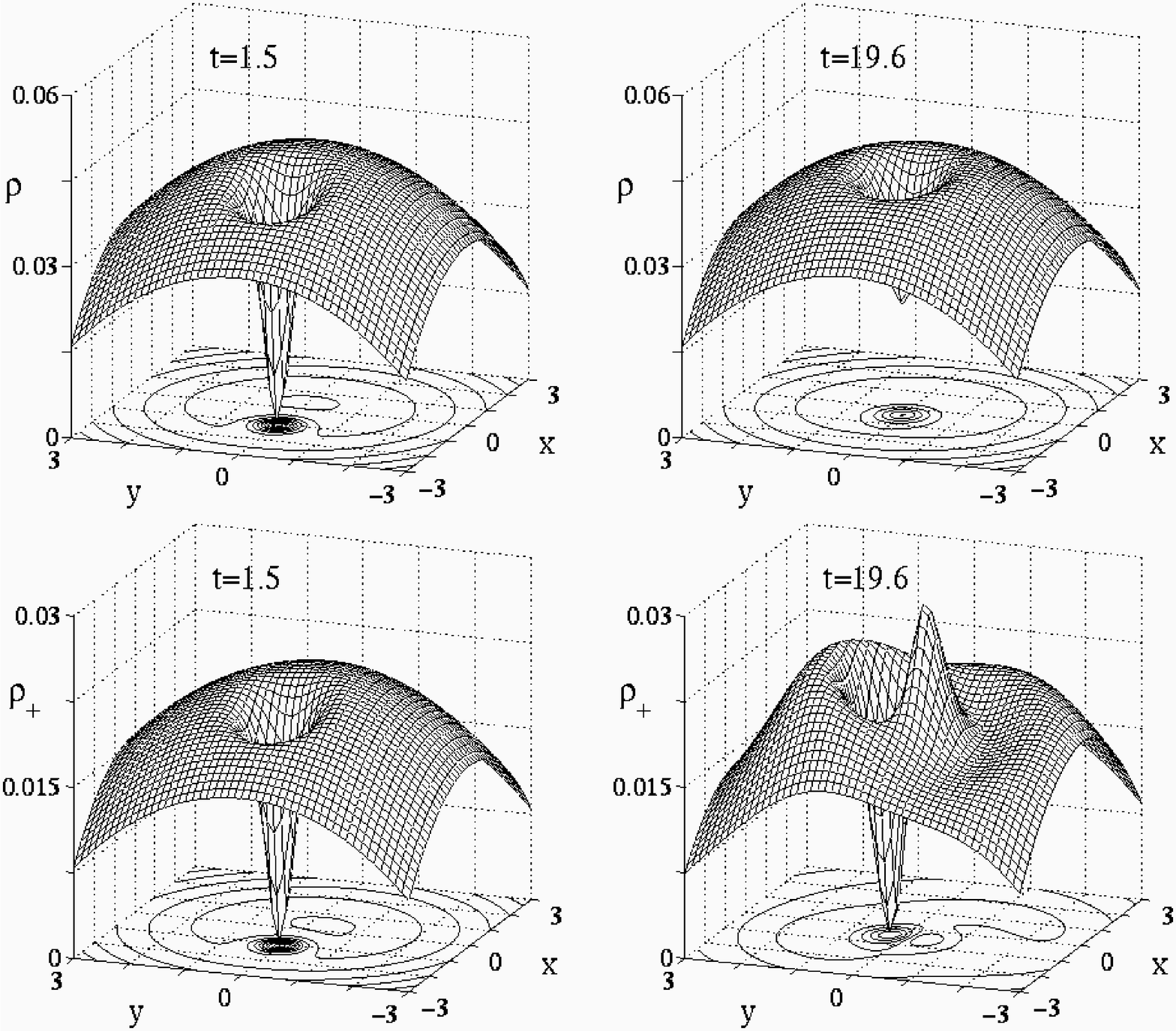, width=.8\linewidth}
 \caption{Monopole instability in the SCV basis.  Top figures show total
density $\rho=|\psi_+|^2 + |\psi_-|^2$ at early and late times; bottom figures
show $|\psi_+|^2$ ($|\psi_-|^2$ being the mirror image in the x-axis.)
No new vortices form; instead, the initially superimposed vortices slip apart,
and each is pinned.}   
\end{figure}
\end{center}

On the other hand, the monopole basis can be more illuminating of
other phenomena.  Two directly aligned monopoles 
perform a curious dance:
they attract each other, merging into a single monopole of winding number
two; this then separates again, but in the perpendicular direction; 
and the cycle 
repeats.  In the SCV basis, this is a rather confusing
system of orbiting vortices, and pinning peaks that form and dissolve; but 
in the monopole basis, it is
quite simple (see Fig.~8).  And the peculiar scattering of monopoles at right
angles in a head-on collision is actually what is expected for
monopoles in gauge theories as well\cite{atiyah}. We find analogous
dances, with $\pi/n$ scattering, for monopoles of higher winding number $n$.

\begin{center}
 \begin{figure}
 \epsfig{file=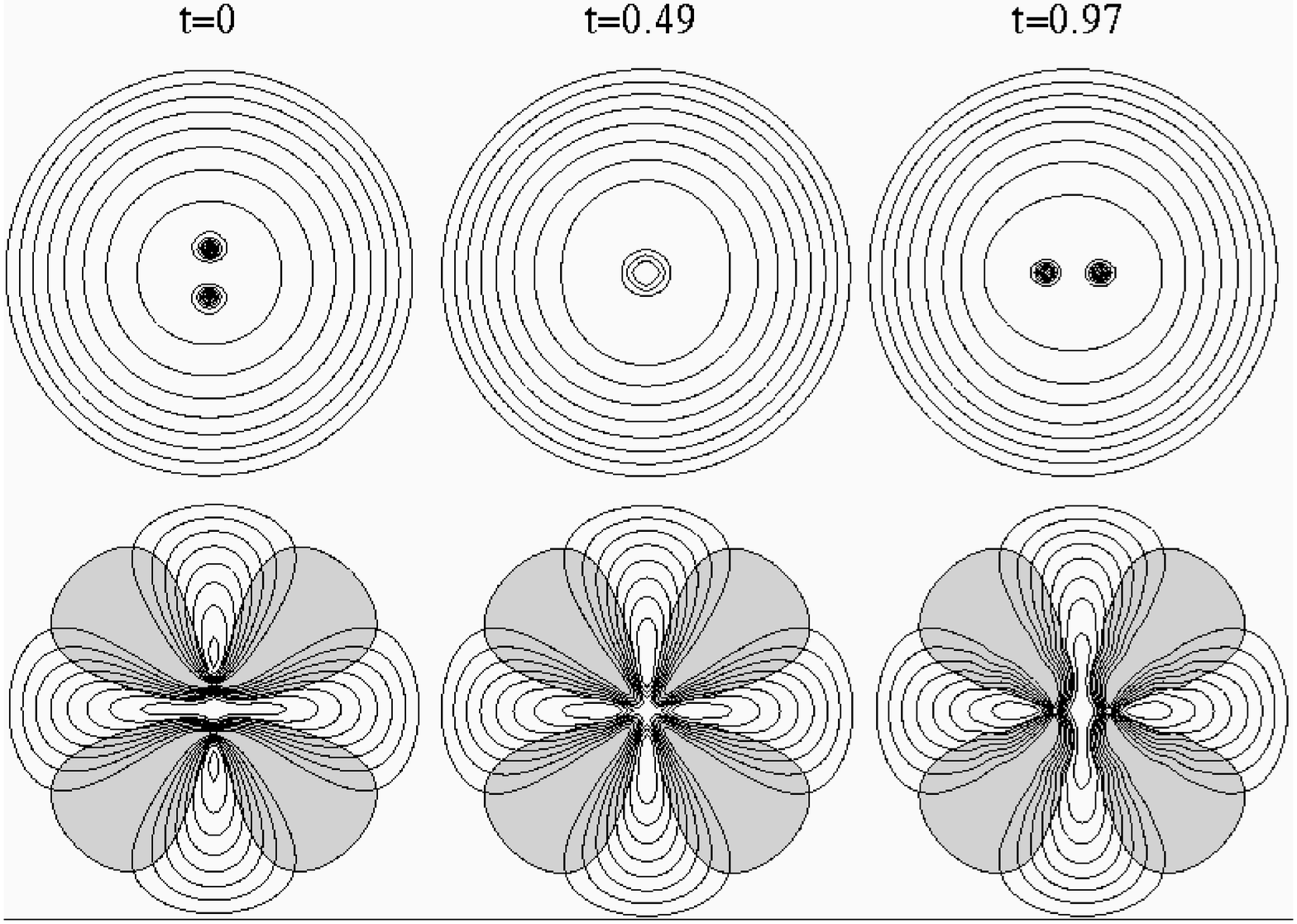,width=.7\linewidth}
 \caption{Two aligned monopoles scattering in a harmonic trap.  
Top figures show total density; bottom figures show 
$|\psi_1|^2$ and $|\psi_2|^2$ separately.
Since $|\psi_2|^2$ essentially does not change, it is merely indicated as 
uniformly shaded lobes.}
 \end{figure}
\end{center}

We conclude our overture to monopoles in Bose condensates
with three remarks on experimental possibilities.  First, despite their
instability, creation of structures like these, in spherical or
toroidal traps, may nevertheless be possible using the adiabatic
passage technique of Dum {\it et al.}\cite{Dum}.  See Fig.~9.
Second, unlike vortices, detection of monopoles should not pose a problem:
since the two species can be imaged separately, the pattern of 
density lobes
is large and obvious, and clearly indicates the presence of a monopole
core.  And finally, the two principles of alternative hydrodynamic
pictures, and using condensates to manipulate condensates, may be
of practical value in a wide range of future experiments on 
multi-component condensates.
\begin{center}
 \begin{figure}
 \epsfig{file=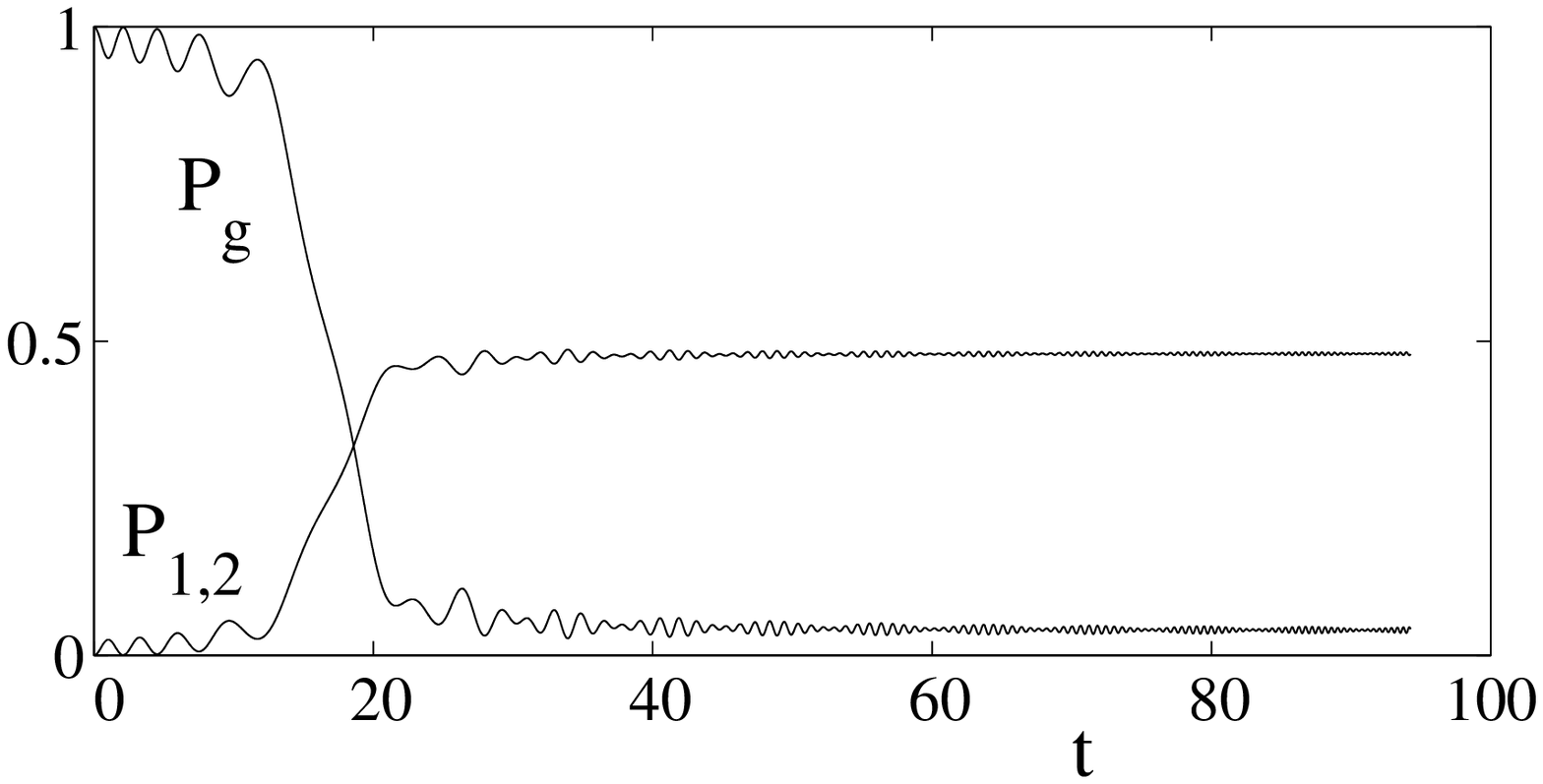, width=.7\linewidth}
\caption{Adiabatic transfer of population (see Ref.~[10]),
from an internal atomic ground state and the motional ground state in the 
trap, to the monopole
configuration of two excited internal states.  We take 
$\lambda_0 k_L=0.1$, $g\int\!d^2x|\psi_j|^2 =1000$, detuning swept linearly 
from -1 to +5 trap frequencies.  
Further parameter optimization may allow total transfer, but there
may also be advantages, for stabilization and observation, to leaving a small
ground state population in the core.}
\end{figure}
\end{center}
We gratefully acknowledge valuable discussions with Ignacio Cirac and 
Peter Zoller, as well as the support of the European Union under the TMR 
Network ERBFMRX-CT96-0002.



\begin{references}
\bibitem{breakdown}  Since mean field theory soon
breaks down in the presence of a dynamical instability,
one should actually quantize these two unstable modes.
\end{references}
\end{document}